# Simulation of the Avalanche Process in the G–APD and Circuitry Analysis of the SiPM

V. M. Grebenyuk, A. I. Kalinin, Nguyen Manh Shat, A.K. Zhanusov, V. A. Bednyakov

Joint Institute for Nuclear Research, Dubna, Russia

## Abstract

The discrete modeling of the Geiger-mode APD is considered. Results of modeling and experimental measurements with the SiPM show that the known formula for the charge of the avalanche pulse $Q=\Delta U * C_d$ underestimates its value. In addition, it is seen from the dynamic of the avalanche multiplication that the resistor $R_q$ in photodiode, usually called a quenching resistor, in reality fulfills only the restoring function. The SiPM restoring time, taken into account the number of pixel N and the load resistance R, is $T=C_d*(R_q+NR)$.

## Introduction

Silicon Photomultipliers (SiPMs) and Geiger photodiodes (G–APDs) [1, 2] are unique instruments capable of detecting individual photons. They are extensively studied and widely used. Some of their properties are probably not adequately studied yet. For example, in all papers dealing with both single G-APD and SiPMs as a whole it is stated that the output charge of one pixel is $Q=E_{OV}*C_d$, where $E_{OV}$ is the excess of the operation voltage over the threshold voltage and $C_d$ is the capacitance of one pixel. It is supposed in this model that the capacitance can be discharged only to this threshold voltage. The same idea underlies the avalanche model proposed by Haintz as far back as 1964 [3] and has been repeated in many papers since that time. However, in a recent paper [4] it was shown by simulation that the output charge is about $2*E_{OV}*C_d$. To describe the diode current and voltage pulse shapes, the authors used discrete simulation of the avalanche process in a Geiger photodiode. Later, a similar doubled result was obtained in [5] by numerical simulation using the same initial formulas as in [4].

In this paper the effect in question is confirmed by simulation [4] in a single G-APD and experimentally in a SiPM.

In addition, the role and the optimum resistance of the quenching resistor in the photodiode are discussed. What one calls a quenching resistor is the resistor placed in series with the diode in each photodiode. It is shown that this resistor does not fulfill the quenching function. In the last section, the equivalent circuit and the shape of the output signal from the silicon photomultiplier are analyzed and some formulas for the SiPM are proposed.

## 2. Simulation of the avalanche process in the Geiger APD.

The discrete model of the photodiode [4] is based on the known phenomenon of ionization by electron and hole impact in a solid. If the photoelectron moving in a high-strength field has a sufficiently high energy, it can ionize atoms of a crystal and thus produce one more electron and one hole. On the other hand, the hole may turn in the same manner into two holes and an electron. The model uses the time-discrete development of the avalanche; one time discreteness step is considered to be development of the avalanche from photoelectrons moving to one side of the diode and then, at the same step, development of the avalanche from the newly arisen holes moving to the other side. Thus, a recurrent avalanche development cycle is considered. A unit of cyclicity is taken to be the time for which the charge carrier passes through the $i$-th layer (multiplication volume). This is favored by the fact that the drift velocity of charge carriers in high-strength fields practically does not change. The duration of one discreteness step is so small that the avalanche process with the introduced cyclicity and discreteness does not significantly differ from the real one. In addition, cyclicity of avalanche model makes it possible to introduce equations in the form of recurrent relations, which simplifies solution of the nonlinear problem.

The photodiode under consideration is based on the n-i-p junction with a narrow volume layer about 1 μm, which is typical of the practically used SiPMs with a sharp p-n junction. Ordinary assumptions are made. The electric field in the multiplication region is uniform. Diffusion and recombination of carriers are ignored. The ionization coefficients for electrons and holes are considered to be known and are taken from the tabulated experimental data: $\alpha(E)$ and $\beta(E)$, On this basis we have an equation for multiplication of electrons per n-i-p junction passage $N_{i+1}(w) = N_i(0) \times 2^{(W*\alpha)}$, where W is the thickness of the avalanche layer and i is the number of the cycle. Half of the carriers are holes, which continue multiplication of particles by factor $0.5W*\beta$ and produce new electrons. Thus, the first equation defines multiplication of carriers per cycle $g=2^{(W\alpha)} \times 0.5W\beta$, and if $g>1$, the development of the process is unlimited, i.e., the number of electrons and holes infinitely increases. The second equation defines a decrease in the capacitance voltage after each cycle (and thus a decrease in g): $U_{i+1}= U_i - [qN_{i+1}(w)-(E_{OV}-U_i) \Delta t/R_q]/C_d$. Hear $C_d$ is the capacitance of the diode and $R_q$ is the resistance placed in series with the diode. When the voltage decreases to the threshold value $U_{br}$, multiplication of the avalanche stops.

Figure 1 shows the results of simulating the photodiode with the sensitive volume 0,7 μm thick and the capacitance 0.25 pF at the working diode voltage $E_{OV}$ = 25 V. The output diode voltage after the launching of the avalanche is shown for five resistors $R_q$=2, 4, 6, 8 and 10 kΩ denoted by the numbers from 1 to 5 respectively.

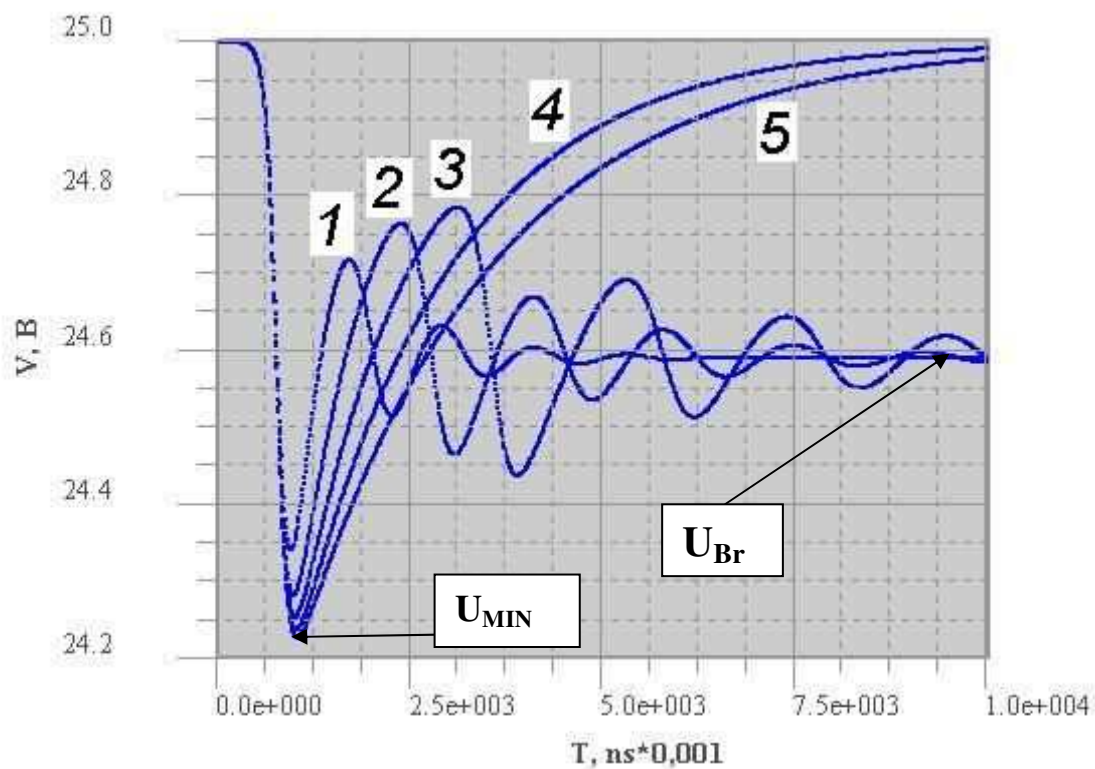

Fig. 1. Photodiode voltage V as a function of time at the operation voltage $E_{OV}$=25 V and diode capacitance 0.25 pF for $R_q$ =2, 4, 6, 8, and 10 kΩ, labeled by numbers 1, 2, 3, 4, and 5 respectively. $U_{br}$ is the threshold voltage.

As the avalanche increases, the diode capacitance is seen to quickly discharge from the initial voltage 25 V to a minimum voltage $U_{min}$. After the avalanche ceases, the voltage at the diode is slowly restored. At the resistance $R_q$ =8 kΩ and higher an exponential increase in the voltage to its initial value is observed. However, at a lower resistance the charging of the capacitance after the cessation of the avalanche stops at the same voltage $U_{br}$ irrespective of the $R_q$ value. Here, the stationary discharging begins instead of the normal pulse discharging. Recurrent launches of the avalanche are observed, but the level of these launches is always higher than $U_{br}$, which confirms the fact that it is exactly the threshold voltage.

To our mind, the picture in Fig. 1 is informative enough to allow qualitative explanation of the avalanche processes occurring in the photodiode. First, the pulse rise time is as small as fractions of a nanosecond. Second, this time, corresponding to the avalanche current pulse duration, is practically independent of the quenching resistor resistance $R_q$. This is because the avalanche current does not pass through this resistor and the avalanche is quenched in the APD by the discharge of the diode capacitance by the intrinsic avalanche current. For example, even at a low resistance $R_q$ = 2 kΩ the avalanche first stops after discharging the diode capacitance, then the diode voltage begins restoring but fails to restore fully, the avalanche is launched again several times, and finally stationary discharging is established. In the normal condition with high resistances the avalanche is launched and quenched only once. Thus, we can reasonably call it a self-quenching mode. Obviously, the avalanche will be happen even at any very high resistance $R_q$, but it is practically an unacceptable because the restoration time will be very large. In the optimum case, the resistor with lowest possible resistance should be chosen for having a high counting rate. This is also necessary because when this resistance decreases, the output current pulse in the silicon photomultiplier increases and the restoring time minimums, though the output charge is constant.

Now we can discuss the question raised at the beginning of the paper. Why does there occur overvoltage $\Delta U = E_{OV} - U_{br}$, in other words, why is the photodiode voltage discharged after launching below the threshold value? To answer the question, let us look at Fig. 1 again. The maximum rise rate corresponding to the current maximum is observed exactly at the threshold voltage irrespective of the resistor's resistance. This means that at this moment multiplication, avalanche increase stops but a large number of electrons and holes are already in the depleted zone and all of them are in motion. Moreover, some of the electrons may even be accelerated, though to a lesser extent. Thus, at the moment of the current pulse maximum electrons have an appreciable kinetic energy, and this energy changes to the potential energy. This manifests itself in that the diode capacitance is discharged by a value larger than $\Delta U \ast C_d$. (Actually, it is an analogue of the ballistic galvanometer used exactly to measure current pulses, in which the maximum overshoot is measured rather than rotation of the galvanometer's coil.) Thus, in the pulsed mode we always observe the voltage overshoot. It is the amplitude of this transient process which is the useful signal. How large is this amplitude?

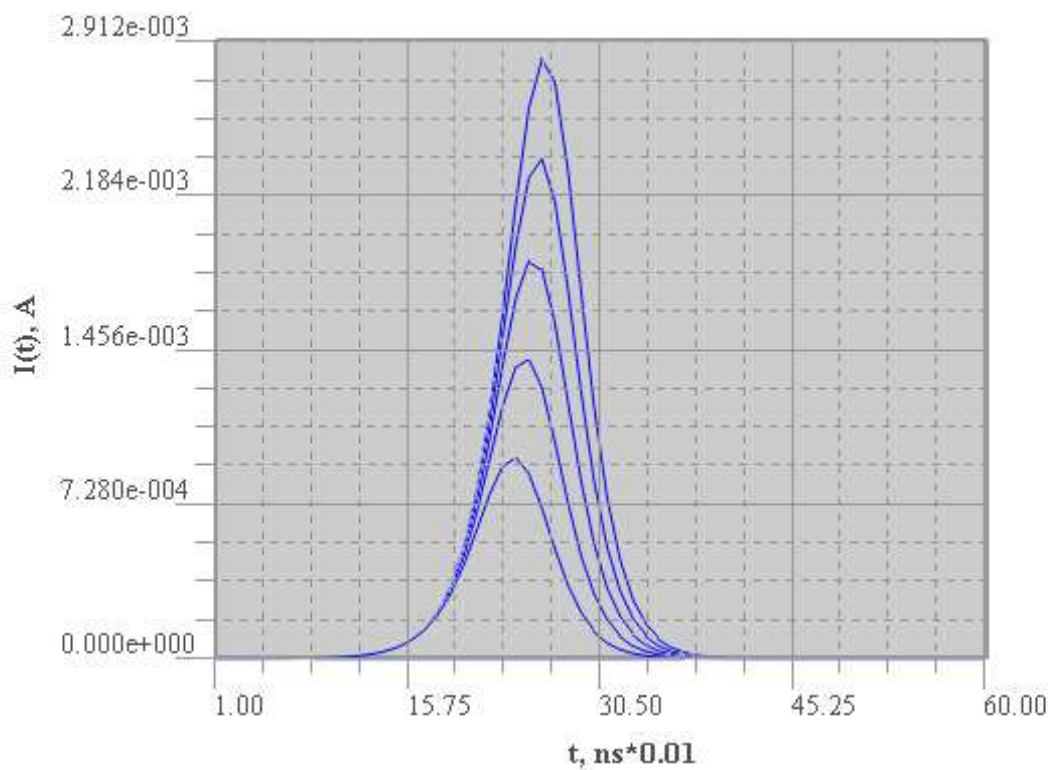

Fig. 2. Shapes of the avalanche current pulses at the diode capacitances 25, 37.5, 50, 62.5 and 75 fF.

It is undoubtedly proportional to $E_{OV}$ and the diode capacitance. The former is evident from secondary launches in Fig. 1. The higher the initial level with respect to the threshold value, the larger the signal amplitude. Figure 2 shows current pulses at different capacitances. The larger the diode capacitance, the larger the current pulse area, i.e., the output charge.

How large is the excess of the pulse amplitude over the threshold $U_{br}$? In Fig. 1, $U_{min}=2 \ast U_{br}$. Simulation at more higher operation voltage shows that the excess can be a factor of 2.5 and larger.

From the point of view of the electrical simulation, this excess of the amplitude is probably governed by the behavior of the ionization coefficients α(E) and β(E) with the field, which shows that they are not zero even below the threshold value. Practically, this excess value is governed by ignored effects. For example, an increase in the diode capacitance with decreasing diode voltage is ignored. However, this may change the amplitude by no more than 10%.

Ignored solid physics phenomena, e.g., electron–electron interaction, may also produce their effect. But this effect limits the maximum current only at very high current densities in the diode, which is unlikely for micrometer transitions.

Direct experimental measurement of the diode pulse amplitude excess is only possible in APD-type diodes, which have the diode–resistor connection output. In SiPMs there is no output of this voltage, and only indirect measurements are possible. The results of this measurement are given in the next section after the consideration of the equivalent SiPM circuit.

## 3. Equivalent SiPM circuit

At the NEC2007 symposium [6] an equivalent circuit of the silicon photomultiplier was proposed for the one-electron event, when one pixel is excited. It was shown that the output signal shape consisted of two exponentially decaying components, the main slow one and the fast but short one. In this paper we extended this circuit to the case where photons excited avalanche processes in n pixels. Figure 3 shows the general circuit of a silicon multiplier, and Fig. 4 shows its extended equivalent circuit. All non-triggered N–n photodiodes are placed in high frequency in parallel with the load resistor, and the triggered diodes are placed in series with it. The voltage pulse with the amplitude $E_{OV}=Q/C_d$ was taken to be the input signal in each particular photodiode.

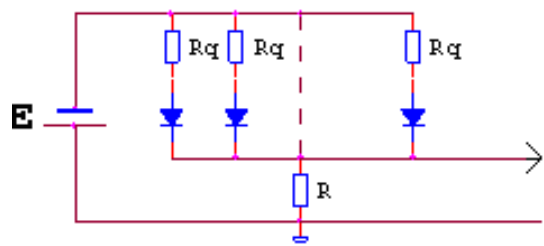

Fig 3. The general equivalent circuit of the SiPM.

Unlike the case in the APD, where the output voltage is picked up at the quenching resistor $R_q$, in the silicon photomultiplier the voltage is picked up at the low-ohmic load resistor R. The formula for the output voltage at the resistor R can be obtained on the basis of the equivalent circuit:

$$U_{sipm} = (nE_{OV}/N) \times R/(R+Z_1/N),$$

where N is the total number of pixels in the SiPM, n is the number of triggered pixels, and $Z_1$ is the total resistance of one pixel which includes $R_q$, $C_q$, and $C_d$.

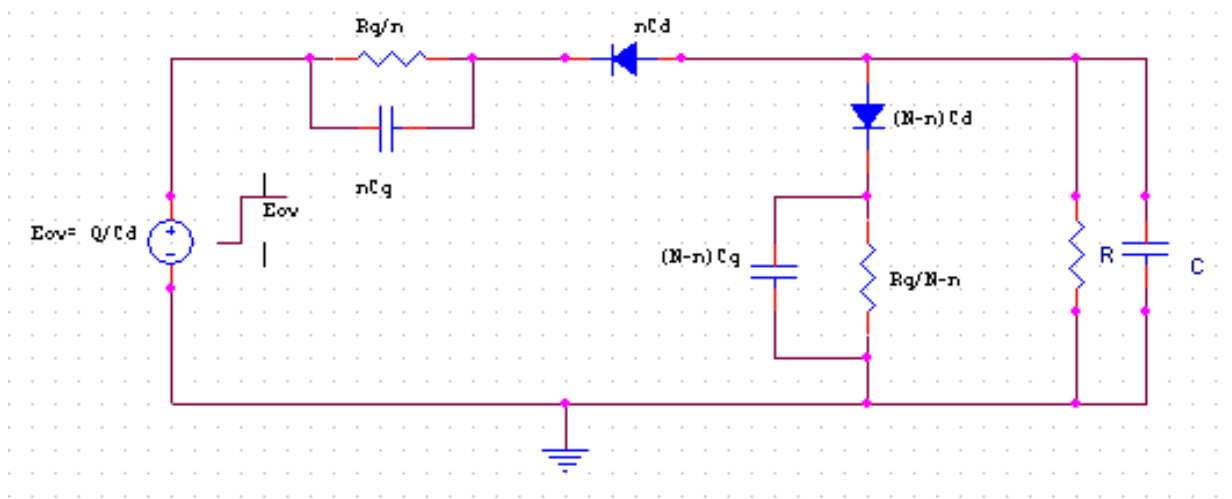

Fig. 4. Equivalent circuit of the SiPM, when n pixels of their total number N were triggered. R and C are the load resistor and capacitor, $R_q$ and $C_q$ are the quenching resistor and the stray-capacitance capacitor placed in parallel with this resistor.

As might be expected, the output signal is proportional to the number of triggered pixels, which is evident from the first term in the formula. The second term shows which part of the avalanche current arrives at the load. This division is frequency dependent and thus governs the shape of the output signal. It is significant that the signal pulse does not depend on the number of triggered pixels but depends on the total number of pixels N. The SiPM output signal shape consists of the exponentially decaying components, the main slow component, whose duration is governed by the time constant $RNC_d + R_q(C_d + C_q)$, and the fast component governed by $R[C_d C_q/(C_d + C_q)]$. As in [6], the rise time of these exponents is taken to be zero. It is allowable because the duration of the avalanche is a few fractions of a nanosecond. It is worth noting that the pulse duration of the SiPM governed by the slow component strongly depends on the load resistance even at R=50 Ω if the number of pixels is very large.

The fast component arises from the passage of the input signal through the capacitance $C_q$ parallel to $R_q$ on its way to the output; therefore, the value of this component depends on the duration of the avalanche.

The two-component shape of the output signal is shown in Fig. 5 [6], which demonstrates oscillograms of two output pulses from a real MP3D SiPM, made by Z.Sadygov.

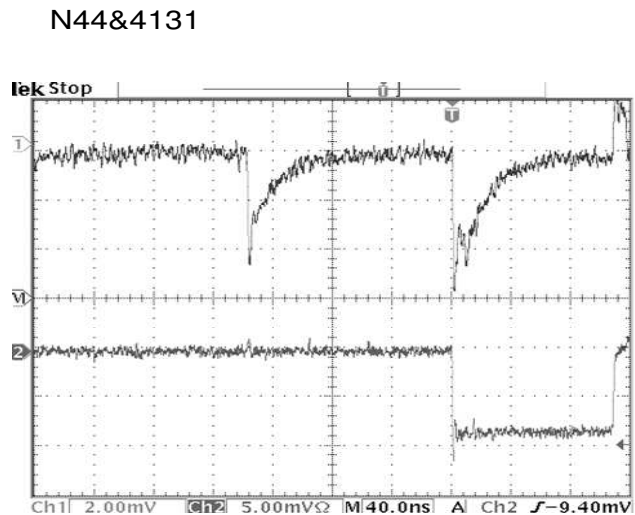

Fig. 5. Oscillograms of two pulses at the SiPM output: the dark single-electron pulse and the generator response pulse (top), and the generator pulse proper (bottom).

The first pulse at the top of the screen is a single-electron noise signal from the SiPM, and the second pulse is from the generator of square pulses supplied to the SiPM input in accordance with the equivalent circuit in Fig. 5. It is seen that the pulses have identical shapes and theirs durations because the pulses are formed in the same circuit.

We have used these oscillograms of the SiPM output signals to evaluate the excess coefficient $K_{ex}$ in the formula for the photodiode output charge $Q_1 = K_{ex} * C_d * E_{OV}$ in the case of a single-electron noise signal, where $E_{OV} = (E_o - U_{brs}) = (44-41) = 3V$ (the diode operation voltage is 44 V, and the threshold voltage is 41 V). In the generator case, $K_{ex} = 1$, but here the input signal passes through all N, connected in parallel, non-triggered diodes, and the amplitude of the signal is measured and is given on the oscillogram $\Delta U_g = 9$ mV. Therefore $Q_g = 1000 * C_d * \Delta U_g$.

Thus, taking into account (from Fig. 5) that $Q_g=1.5*Q_1$, one can easy find the excess coefficient, $K_{ex}=9mV*1000*C_d/3V*1.5*C_d=2$, and therefore $Q_1 = 2*C_d*\Delta U$.

## Conclusions

1. The discrete simulation of the Geiger avalanche photodiode has shown that the main avalanche quenching reason is not the resistor $R_q$ placed in series with the diode but the discharge of the diode capacitance by the internal avalanche current.
2. By simulation and experimentally it is confirmed that it is possible to get the output charge of SiPM about twice as large as $Q=C_d*\Delta U$.
3. The discrete model of avalanche development in photodiodes, which was proposed in [3], makes it possible to get many important parameters of the photodiode: the current pulse shape, the capacitor voltage, their dependence on the diode capacitance, on the diode operation voltage, etc.
4. The circuitry analysis of the SiPM shows that at a large number of pixels the pulse duration of the photodiode increases and overlap of events is possible even at a diode load as small as, for example, 50 Ω.

The authors are grateful to A.G. Olshevsky for his interest and support of the work.

## References

1. P.Buzhan, B.Dolgoshein, L.Filatov et al. Large area silicon photomultipliers: Performance and applications. NIM A, 567 (2006) 76.
2. Z.Sadygov, A.Olshevski, I.Chirikov et al. Three advanced designs of micro-plxel avalanche photodiodes: their present maximum possibilities and limitatios. NIM A, 567 (2006) 70.
3. R.H. Haintz. Model for the electrical behavior of a microplasma. J. Appl. Phys. 35, 1370, 1964.
4. I.V.Vanjushin, V.A.Gergel, V.M.Gontar et al. Physics and Technique of semicondactors, vol. 41, issue 6, 2007.
5. H.Otono, H.Oide, S.Yamashita, T.Yoshika. arXiv:0808.2546
6. D.Chokheli, V.Grebenyuk, A.Kalinin. Study of the SiPM signal shape with the help of transient functions. NEC2007, Proceedings of the 21st International Symposium, Varna, Bulgaria/